\documentstyle[aps,multicol,pre,psfig,epsf]{revtex}
\begin{document}
\title
{
Solitons in the
noisy Burgers equation
}
\author{Hans C. Fogedby and Axel Brandenburg}
\address
{
Institute of Physics and Astronomy,
University of Aarhus, DK-8000, Aarhus C, Denmark
\\
NORDITA, Blegdamsvej 17, DK-2100, Copenhagen {\O}, Denmark
}
\date{\today}
\maketitle
\begin{abstract}
We investigate numerically the coupled diffusion-advective type 
field equations
originating from the canonical phase space approach to
the noisy Burgers equation or the equivalent Kardar-Parisi-Zhang
equation in one spatial dimension. The equations support
stable right hand and left hand solitons and in the low
viscosity limit a long-lived
soliton pair excitation. We find that two identical pair excitations
scatter transparently subject to a size dependent phase shift and that
identical solitons scatter on a static soliton transparently without a
phase shift. The soliton pair excitation and the scattering 
configurations are interpreted in terms of growing step and 
nucleation events in the interface growth profile.
In the asymmetrical case the soliton  
scattering modes are unstable presumably toward multi soliton
production and extended diffusive modes, signalling the general
non-integrability of the coupled field equations.
Finally, we have shown that growing steps perform anomalous
random walk with dynamic exponent $z=3/2$ and that the nucleation
of a tip is stochastically suppressed with respect to plateau
formation.
\end{abstract}
\draft
\begin{multicols}{2}
\section{Introduction}
There is a continuing interest in the strong coupling aspects
of stochastically driven nonequilibrium systems.
The phenomena in question are ubiquitous and comprise
turbulence in fluids, interface and growth problems, and chemical
and biological systems.

In this context the noisy Burgers equation or the equivalent
Kardar-Parisi-Zhang (KPZ) equation, describing the nonequilibrium growth
of a noise-driven interface, provide a simple continuum
model
of an open driven nonlinear system exhibiting scaling and pattern
formation.

In one dimension, which is our concern here, the noisy Burgers 
equation for the local slope,
$u(x,t)=\nabla h(x,t)$, of a growing interface has the form
\cite{Forster76,Forster77}
\begin{eqnarray}
&&\frac{\partial u}{\partial t}=
\nu\nabla^2u + \lambda u\nabla u + \nabla\eta ~, 
\label{bur}
\\
&&\langle\eta(xt)\eta(00)\rangle = \Delta\delta(x)\delta(t) ~.
\label{noise}
\end{eqnarray}
The height profile (in a comoving frame) 
$h(x,t)$ is then governed by 
the equivalent KPZ equation 
\cite{Kardar86,Medina89}
\begin{eqnarray}
\frac{\partial h}{\partial t} = 
\nu\nabla^2 h + \frac{\lambda}{2}(\nabla h)^2 + \eta ~.
\label{kpz}
\end{eqnarray}
In (\ref{bur}) and (\ref{kpz}) $\nu$ is the damping or viscosity
characterizing the 
linear diffusive term, $\lambda$ a coupling strength for the nonlinear
mode coupling or growth term, and $\eta$ a Gaussian 
white noise,
driving the system into a statistically stationary state. The noise is correlated
according 
to
(\ref{noise}) and characterized by the strength $\Delta$. Moreover, the Burgers
equation is invariant under the slope-dependent Galilean
transformation
\begin{eqnarray}
x\rightarrow x - \lambda u_0t ~, ~~~~u\rightarrow u + u_0 ~,
\label{gal}
\end{eqnarray}
i.e., the interface is superimposed with a constant slope in a moving frame.

The Burgers equation (\ref{bur}) and its KPZ equivalent in one and
higher dimensions have been the subject of intense scrutiny in recent
years
owing to its paradigmatic significance within the field theory 
of nonequilibrium
systems
\cite{Halpin95,Barabasi95,Krug97,Frey94,Frey96,Frey99,Janssen99,Laessig98a,Laessig00,Colaiori01a,Colaiori01b}.

In a series of papers the one dimensional case
defined by (\ref{bur}) and (\ref{noise}) 
has been analyzed in an attempt to uncover
the
physical mechanisms underlying the pattern formation and scaling
behavior.
Emphasizing that the noise strength $\Delta$ constitutes the  relevant 
nonperturbative parameter that is driving the system into a statistically stationary state,
the method was initially based on a weak noise saddle point
approximation to the Martin-Siggia-Rose functional formulation
of the
noisy Burgers equation \cite{Martin73,Baussch76,Fogedby98b,Fogedby98c}.
This work was a continuation of earlier work based on the mapping
of a solid-on-solid model onto a continuum spin model
\cite{Fogedby95}.
More recently the functional approach has been
superseded by a {\em canonical phase space method} 
deriving from the 
canonical structure of the Fokker-Planck 
equation associated with the
Burgers equation
\cite{Freidlin84,Graham84,Graham89,Fogedby99a,Fogedby99b}.
Below we briefly summarize these findings.

The functional or the equivalent phase space approach valid in the
weak noise limit, $\Delta\rightarrow 0$, replaces the stochastic
Langevin-type Burgers equation (\ref{bur}) by coupled deterministic
diffusion-advection type mean field equations,
\begin{eqnarray}
\frac{\partial u}{\partial t}&&=
\nu\nabla^2 u -\nabla^2p+\lambda u\nabla u ~,
\label{mfe1}
\\
\frac{\partial p}{\partial t}&&=
-\nu\nabla^2 p +\lambda u\nabla p ~,
\label{mfe2}
\end{eqnarray}
for the slope $u(x,t)$ and a canonically conjugate noise field
$p(x,t)$, replacing the stochastic noise $\eta$. The field equations
bear the same relation to the Fokker-Planck equations as the classical
equations of motion bear to the Schr\"{o}dinger equation in the 
semi-classical WKB approximation \cite{Landau59c}.

To justify the weak noise limit we recall the
analogy with the WKB approximation in quantum mechanics which, owing to
its nonperturbative character, captures features like 
bound states and tunneling amplitudes, which are generally inaccessible
to perturbation theory. Therefore, we anticipate that the present
weak noise approach to the Burgers equation also accounts 
correctly, at least in a qualitative sense, for the stochastic properties
even at larger noise strength. However, there may be an upper
threshold value beyond which the
system may enter a new stochastic or kinetic
phase. In the one dimensional case discussed here
the scaling behavior is controlled by a single
strong coupling fixed point which can be accessed
by the present weak noise approach. In two and higher
dimension a dynamic renormalization group analysis
predicts a kinetic phase transition at a critical noise
strength (or coupling strength) and the weak noise
approach presumably fails. 

The equations (\ref{mfe1}) and (\ref{mfe2}) derive from
a principle of least action characterized 
by an action $S(u'\rightarrow u'',t)$ associated with an 
orbit $u'(x)\rightarrow u''(x)$ traversed in time $t$ \cite{Landau59b},
\begin{eqnarray}
S(u'\rightarrow u'',t) = \int_{0,u'}^{t,u''}dt\,dx
\left(p\frac{\partial u}{\partial t} - {\cal H}\right) ~,
\label{act}
\end{eqnarray}
with Hamiltonian density
\begin{eqnarray}
{\cal H} = p\left(\nu\nabla^2u + \lambda u\nabla u - 
\frac{1}{2}\nabla^2 p\right) ~.
\label{hamden}
\end{eqnarray}
The action is of central importance in the present approach
and serves as a  weight function for the 
noise-driven
nonequilibrium configurations in much the same
manner as the energy $E$ in the Boltzmann factor $\exp(-\beta E)$ for
equilibrium systems, where $\beta$ is the inverse temperature. 
The dynamical action in fact
replaces the energy in the context of the dynamics of stochastic 
nonequilibrium systems governed by a generic Langevin equation
driven by Gaussian white noise.
The action provides a methodological approach and 
yields access to the
time dependent and stationary probability distributions,
\begin{eqnarray}
&&P(u'\rightarrow u'',t)\propto
\exp\left[-\frac{S(u'\rightarrow u'',t)}{\Delta}\right] ~,
\label{dis} 
\\
&&P_{\text{st}}(u'') = \lim_{t\rightarrow\infty}P(u'\rightarrow u'',t) ~,
\label{statdis}
\end{eqnarray}
and associated moments, e.g., the stationary slope correlations
\begin{eqnarray}
&&\langle u(xt)u(00)\rangle = 
\nonumber
\\
&&\int\prod du~ u''(x)u'(0)P(u'\rightarrow u'',t)P_{\text{st}}(u') ~.
\label{cor}
\end{eqnarray}

The canonical formulation associates the conserved energy $E$
(following from time translation invariance), the conserved
momentum $\Pi$ (from space translation invariance), and the 
conserved area $M$ (from the Burgers equation with conserved
noise):
\begin{eqnarray}
&&E=\int dx\,{\cal H}~,
\label{ener}
\\
&&\Pi = \int dx\, u\nabla p~,
\label{mom}
\\
&&M = \int dx\, u~.
\label{area}
\end{eqnarray}

The field equations (\ref{mfe1}) and (\ref{mfe2}) determine orbits
in a canonical $up$ phase space where the dynamical issue in
determining $S$ and thus $P$ is to find an orbit from $u'$ to
$u''$ in time $t$, $p$ being a slaved variable. Note that unlike
dynamical system theory we are not considering the asymptotic
properties of a given orbit. In general 
the orbits in phase space lie on the manifolds
determined by the constants of motion $E$, $\Pi$ and $M$. Here the
zero energy manifold $E=0$ plays a special role in defining the
stationary state. For vanishing or periodic boundary conditions
for the slope field the zero energy manifold is  
composed of the
transient submanifolds $p=0$ and the stationary submanifold 
$p=2\nu u$.
The zero energy
orbits on the $p=0$ manifold correspond to solutions of the
damped noiseless Burgers equation; the orbits on the $p=2\nu u$
are solutions to the undamped noiseless Burgers equation with negative
damping, i.e., $\nu$ replaced by $-\nu$. In the solvable linear case of the
noise driven diffusion equation for $\lambda=0$, i.e., the Edwards-Wilkinson
equation \cite{Edwards82}, a finite 
energy orbit from $u'\rightarrow u''$ in time $t$
migrates to the zero energy manifold in the limit $t\rightarrow\infty$,
yielding according to Eqs. (\ref{act}) and (\ref{statdis}) the stationary
distribution $P_{\text{st}}\propto\exp(-(\nu/\Delta)\int dx\, u^2)$.
This distribution also holds in the Burgers case and is a generic result
independent of $\lambda$ \cite{Huse85}. Finally, in the long time limit an orbit
from $u'\rightarrow u''$ is attracted to the hyperbolic saddle point
at the origin in phase space implying ergodic behavior in the stationary
state. In Fig.~1 we have schematically depicted possible orbits in phase
space.

The field equations (\ref{mfe1}) and (\ref{mfe2}) admit nonlinear
soliton or smoothed shock wave solutions which are, in the static case, of
kink-like form,
\begin{eqnarray}
u_1(x) = 
u\tanh\left[\frac{\lambda|u|}{2\nu}x\right] ~.
\label{sol}
\end{eqnarray}
Propagating solitons are subsequently generated by the Galilean boost 
(\ref{gal}).
Denoting the right and left boundary values by $u_+$ and $u_-$,
respectively,
the propagation
velocity is given by
\begin{eqnarray}
u_+ +  u_- = -2v/\lambda ~.
\label{solcon}
\end{eqnarray}
\begin{figure}
\begin{picture}(100,170)
\put(-20.0,20.0)
{
\centerline
{
\epsfxsize=8cm
\epsfbox{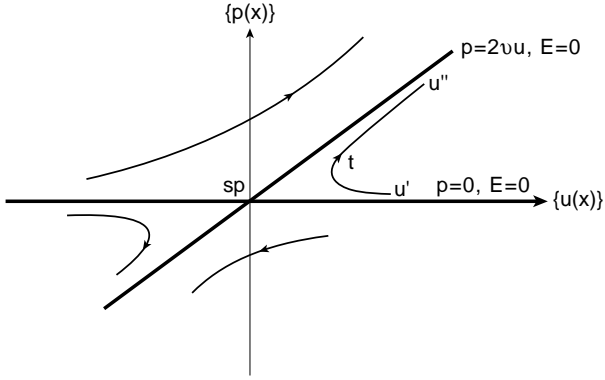}
}
}
\end{picture}
\caption
{
Generic behavior of the orbits in $up$ phase space.
Heavy lines indicate the zero energy manifold. The stationary
saddle point (sp) is at the origin. The finite time orbit from
$u'$ to $u''$ is attracted to the saddle point for $t\rightarrow\infty$.
}
\end{figure}

The amplitude of the static soliton is $u$ and the soliton
is located at the origin.
The right hand soliton for $u>0$, i.e., the soliton with the
larger right hand side boundary value, moves on the
noiseless manifold $p=0$ and is also a solution of the damped
(stable) noiseless
Burgers equation for $\eta =0$. The noise-induced left hand
soliton for $u<0$, i.e.,  the soliton with the larger left hand side
boundary value, is associated with the noisy manifold $p=2\nu u$, 
and is a solution of the undamped
(unstable) noiseless Burgers equation with $\nu$ replaced by
$-\nu$. 
In addition the field equations also 
admit linear mode solutions superimposed 
as ripple modes on the solitons \cite{Fogedby00b}.
The ripple modes are superpositions of both decaying and growing
components reflecting the noiseless and noisy manifolds $p=0$ and
$p=2\nu u$, respectively.
The soliton mode induces a propagating component with
velocity $\lambda u$ in such a way that the right hand soliton acts
like a sink and the left hand soliton as a source of linear modes.
In the Edwards-Wilkinson limit for $\lambda\rightarrow 0$ the ripple
modes become the usual diffusive modes (growing and decaying) of the
driven stationary diffusion equation.

The heuristic physical picture that emerges from our analysis is that of a 
many body formulation
of the pattern formation of a growing interface in terms of a dilute
gas of propagating solitons matched according to the soliton condition
(\ref{solcon}) with superimposed linear ripple modes. In Fig.~2 we have depicted
the basic soliton modes constituting the building blocks in the 
representation of a growing interface. For further illustration we have
shown in Fig.~3 the slope field $u$, the corresponding height field $h$,
and the noise field $p$ for a 4-soliton configuration.

In the present paper we embark on a numerical analysis of the coupled
field equations (\ref{mfe1}) and (\ref{mfe2}) with the purpose of 
investigating them in more detail and provide a numerical underpinning
of the heuristic quasi-particle picture
advanced in the work referred to above. The paper is organized in the
following manner. In Section~2 we discuss the soliton modes.
In Section~3 we introduce the numerical
method designed in order to treat the inherent instability.
In Section~4 we present our numerical
results for the scattering of two single solitons on a static soliton
and the scattering of two  soliton pairs.
In Section~5 we discuss growth and nucleation associated with the
modes investigated numerically. 
Section~6 is devoted to a summary of our 
results and a conclusion.
\begin{figure}
\begin{picture}(100,145)
\put(-10.0,10.0)
{
\centerline
{
\epsfxsize=8cm
\epsfbox{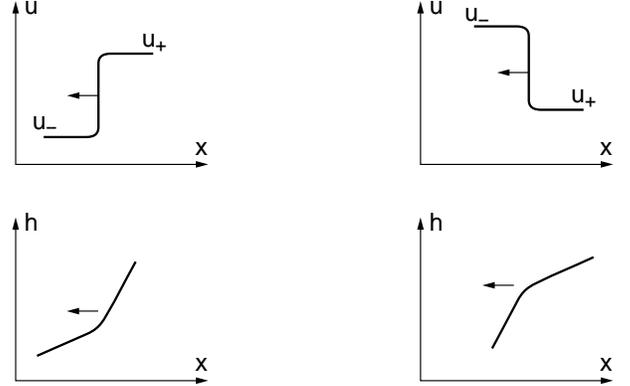}
}
}
\end{picture}
\caption{
Slope field $u$ and height profile $h$ for
the right hand and left hand moving kink solitons forming
the `quarks' in the description of a growing interface.
}
\end{figure}
\begin{figure}
\begin{picture}(100,320)
\put(0.0,20.0)
{
\centerline
{
\epsfxsize=6cm
\epsfbox{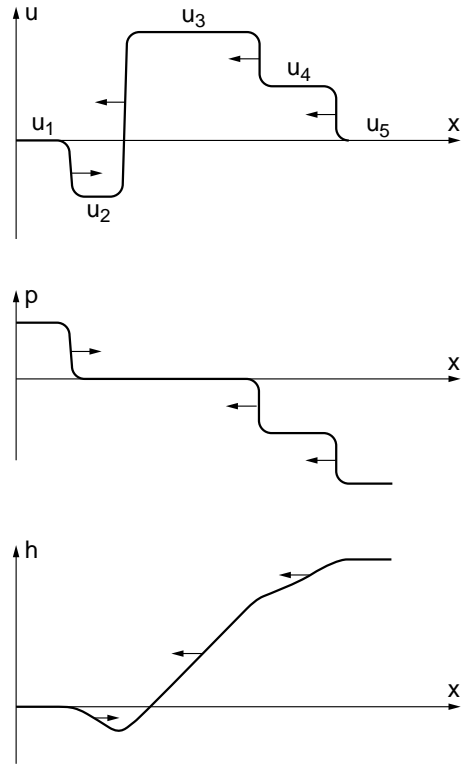}
}
}
\end{picture}
\caption{
4-soliton representation of the slope field
$u$, the noise field $p$, and the height field $h$.
}
\end{figure}

\section{Soliton modes}
The exact right and left hand soliton solutions of the field
equations do not satisfy periodic or vanishing boundary
conditions in the slope field $u$; the nonvanishing boundary
values $u_+$ and $u_-$ in fact correspond to a deterministic current 
dissipated or generated at the soliton centers yielding 
permanent
profile solutions \cite{Fogedby98a}. The kink solitons 
constitute the elementary
building blocks or `quarks' in the present approach and the interface 
profile is then built up by matching solitons according to the
matching condition (\ref{solcon}).

The simplest mode satisfying periodic boundary conditions is the
two-soliton or pair soliton configuration
\begin{eqnarray}
u_2(x,t)=u_1(x-vt-x_1) - u_1(x-vt-x_2)~,
\label{2sol}
\end{eqnarray}
obtained by matching a right hand and a left hand soliton boosted to the
velocity $v=-\lambda u$. The two-soliton mode has amplitude $2u$ and
size $|x_2-x_1|$. The associated noise field vanishes for the right
hand component and equals $2\nu u$ for the left hand component; 
we thus have
\begin{eqnarray}
&&p_2(x,t) = -2\nu u_1(x-vt-x_2)
~~\text{for}~~ u>0~,
\label{2p1}
\\
&&p_2(x,t) = +2\nu u_1(x-vt-x_1)
~~\text{for}~~ u<0~.
\label{2p2}
\end{eqnarray}
By inspection it is seen that the pair mode (\ref{2sol}) is an
approximate solution to the field equations (\ref{mfe1}) and
(\ref{mfe2}). The correction terms are of the type $u\nabla u$ and
$u\nabla p$ referring to the distinct components of $u_2$ and $p_2$
and thus correspond to local perturbations from a region of size
$\nu/\lambda|u|$ which is small in the low viscosity limit
$\nu\rightarrow 0$. We assume that the correction can be treated within
a linear stability analysis and thus gives rise to a linear mode
propagating between the 
right hand and left hand solitons \cite{Fogedby00b}.

The pair mode thus forms a long-lived excitation or quasi-particle 
in the many body description of a growing interface. Subject to periodic
boundary conditions this mode corresponds to a simple growth situation.
The propagation of the pair mode corresponds to the propagation of a
step in the height field $h$. At each revolution of the pair mode the
interface grows by a uniform layer of thickness $2u|x_2-x_1|$. In Fig.~4
we have depicted the pair mode in $u$, the associated noise field $p$,
and the height profile $h$.

Generally a growing interface, ignoring the superimposed linear
ripple modes, can at a given time instant be re\-pre\-sented by a
gas of matched left hand and right hand solitons as depicted in
Fig.~3 in the 4-soliton case. A gas of pair solitons thus constitute
a particular growth mode where the height profile between moving
steps has horizontal segments.

\begin{figure}
\begin{picture}(100,280)
\put(0.0,10.0)
{
\centerline
{
\epsfxsize=7cm
\epsfbox{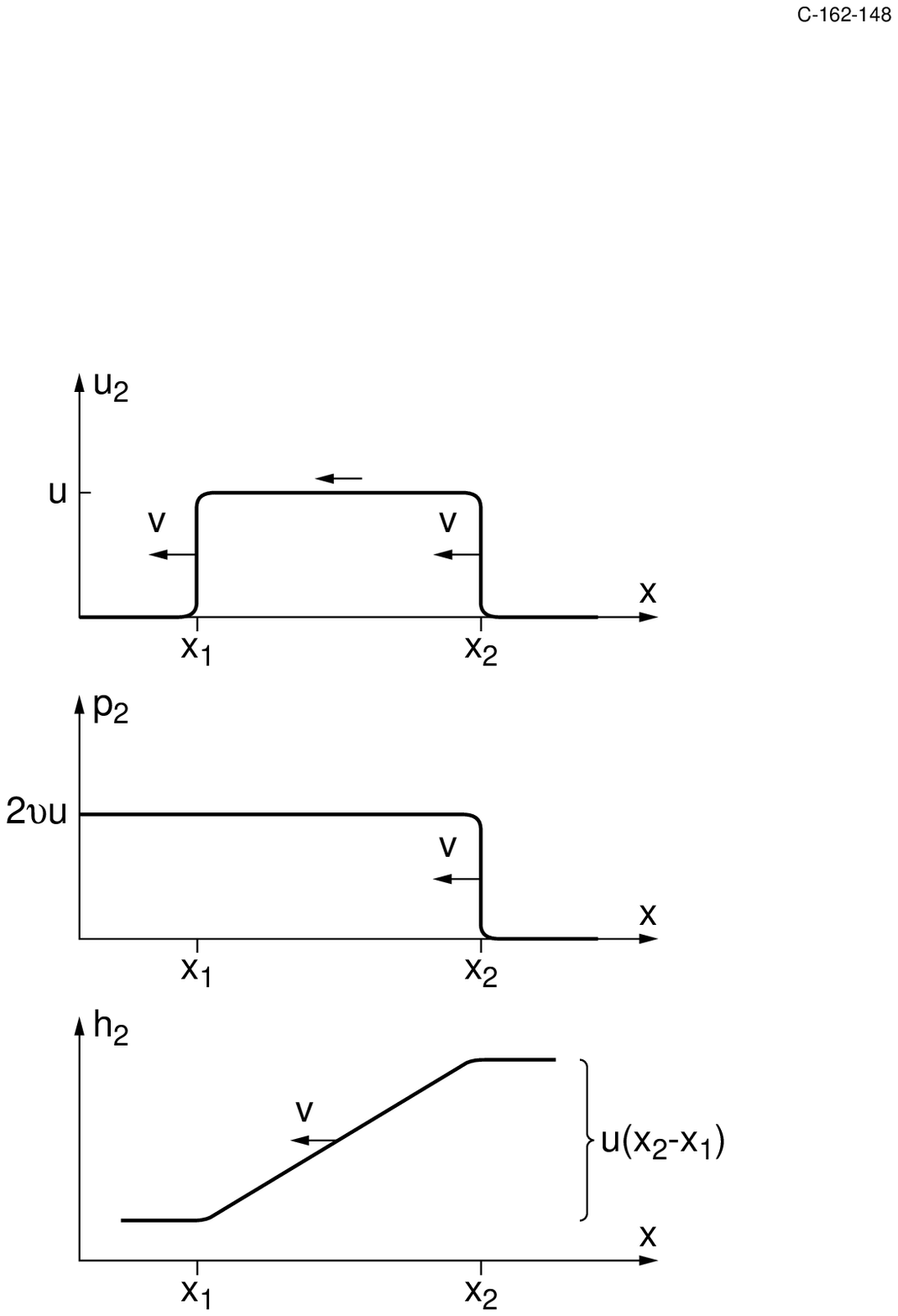}
}
}
\end{picture}
\caption{
Slope field $u_2$, the associated noise field
$p_2$, and the resulting height profile $h_2$ at time $t=0$
for a two-soliton configuration.
}
\end{figure}

The soliton picture also allows us easily to understand in what sense
the right hand soliton acts like a drain and the left hand soliton
as a source with respect to perturbations. Considering two pair solitons
superimposed on the right and left horizontal parts of the static solitons
(\ref{sol}) it follows from (\ref{solcon}) that for a right hand
soliton perturbations move toward the soliton center
and for a left hand soliton perturbations move
away from the soliton center. This mechanism also follows from the
linear ana\-ly\-sis of ripple modes in \cite{Fogedby00b}. The mechanism is depicted
in Fig.~5
\begin{figure}
\begin{picture}(100,100)
\put(-10.0,20.0)
{
\centerline
{
\epsfxsize=8cm
\epsfbox{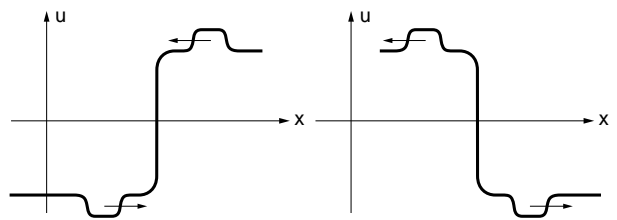}
}
}
\end{picture}
\caption{
Source and drain mechanism for the right hand and left
hand solitons. The perturbation attracted and repelled by the
soliton centers are modeled by pair solitons.
}
\end{figure}

Since we do not posses explicit propagating 
multi-soliton solutions of the field equations
(\ref{mfe1}) and (\ref{mfe2}) the problem of soliton
collisions remains unresolved from an analytical
point of view. Therefore we now turn to
a numerical analysis of the problem.
\section{Numerical method}
The coupled field equations (\ref{mfe1}) and (\ref{mfe2}) are
of the diffusion-advection type with the exception that the evolution of
$p$ is governed by
a negative diffusion coefficient. Standard numerical methods designed 
to step the equations
forward in time fail because small perturbations with wavenumber $k$
grow in time like $\exp(\nu k^2 t)$, so perturbations
with the largest possible $k$ grow fastest, rendering hence the integration
unstable. In order to circumvent this problem we have designed a method
to solve the equations iteratively starting with a trial solution for
$u$ and $p$ in the space interval $|x|\leq L$ and
time interval $0\leq t\leq T$. At each iteration step we proceed in two sweeps.
In the forward sweep we step only the equation for $u$ forward in time
using $p(x,t)$ from the previous iteration step. In the backward
sweep we step the equation for $p$  backward in time, using $u(x,t)$ from
the forward sweep. In this manner the unwanted perturbations in $p$
decrease exponentially as one moves backward in time.

A drawback of this method is that we can specify initial values only
for $u$, and that we must instead specify $p$ at the final time $t=T$. 
In the present
context where we want to consider collisions of solitons this is not
a serious problem, because here it is possible to guess the final
solution. The numerical solution serves therefore mainly as a tool
to check that a certain guess is actually a solution. Furthermore,
this method allows us to calculate the precise functional form
of $u$ and $p$ during the collision, even if the initial guess around
the time of the collision was actually wrong.

We have solved the equations on a mesh with $N_x=1001$ mesh points
and $N_t=6001$ time steps. For $L=1/2$ we have a mesh spacing
of $\Delta x=0.001$. For both sweeps we use sixth order finite
differences to calculate first and second derivatives and a third
order Runge-Kutta scheme for the time integration (see, e.g.,
the appendix of Ref.~\cite{Sanchez01} for these schemes).

In order to adequately resolve $u$ and $p$
at all times we must choose a suitable value of $\nu$. We found
empirically that $\nu=0.005$ gave good results, which is the value
adopted
in the following. For smaller values of $\nu$ the $u$ and $p$ functions
become only marginally resolved whereas for larger values of $\nu$
the length of the time step is mostly controlled by the value of $\nu$
rather than just the propagation speeds of the solitons.
Empirically we found that the maximum time step that
can be used is $\Delta t=5\times10^{-5}$ for $\nu=0.005$. 
In all cases we have chosen $\lambda=1$.
\section{Solitons, Pair solitons, and soliton collisions}
In choosing soliton configurations to be verified by the numerical
method we have found that it is essential to satisfy the three
conservation laws governing the dynamics of solitons, namely the 
conservation of energy (\ref{ener}), the conservation of momentum
(\ref{mom}), and the conservation of area (\ref{area}).
\subsection{Solitons and pair soliton}
We have numerically verified that the right and left hand solitons
(\ref{sol}) for $u>0$ and $u<0$ are solutions. By construction the 
two-soliton configuration (\ref{2sol}) carries energy 
$E_2 = (-16/3)\lambda\nu|u|^3$, momentum $\Pi_2=-4\nu u|u|$, and
for small $\nu$ the area $M_2\propto 2\nu|x_1-x_2|$. We have
shown that, for small $\nu$, a well-separated two-soliton mode, i.e.\ with
$|x_1-x_2|\gg\nu/\lambda u$, is a long-lived
excitation, hence justifying the heuristic argument.

In order to lend support to the heuristic quasi-particle picture
based on the elementary kink solitons (quarks) and the composite
pair soliton as the basic quasi-particle it is essential to consider
soliton collisions. We have here considered the two symmetric cases:
i) the collision of two propagating solitons with a static soliton
and ii) the collision of two pair solitons. In both cases the
configurations are symmetric and the conservation laws are satisfied at all
times including the collision regime.
\subsection{Three soliton collisions}
In the first case two propagating solitons moving in opposite
directions collide with a static soliton located at the center.
The trial solution has the form
\begin{eqnarray}
&&
u(x,t) = -\text{sign}(t)[u_1(x+vt)+u_1(x-vt)-2u_1(2x)]
\nonumber
\\
&&
\label{3u}
\\
&&
p(x,t) = -2\nu[u_1(x+vt)+u_1(x-vt)]
~~\text{for}~~t<0
\label{3p1}
\\
&&
p(x,t) = -4\nu u_1(2x)
~~\text{for}~~t>0
\label{3p2}
\end{eqnarray}
and the height field
\begin{eqnarray}
&&h(x,t) = -\text{sign}(t)\left[\frac{2\nu}{\lambda}\right]
\times
\nonumber
\\
&&\log
\left|\frac
{
\cosh\left[\frac{\lambda|u|}{2\nu}(x-vt)\right]
\cosh\left[\frac{\lambda|u|}{2\nu}(x+vt)\right]
}
{
\cosh\left[\frac{\lambda|u|}{2\nu}x\right]
}
\right|
\label{3h}
\end{eqnarray}
with velocity $v=\lambda u$.

In this mode two left hand solitons with amplitude $2u$ propagate
with equal and opposite velocities toward a static right hand soliton
with amplitude $4u$ located at the center. During the collision the left
hand solitons are absorbed, the static right hand soliton flips over
to a static left hand soliton and two right hand solitons emerge
propagating away from the center with equal and opposite velocity.
The solitons thus collide transparently with the static soliton, i.e.,
there is no reflection, and there is no phase shift associated 
with the collision.
In terms of the associated height profile this scattering situation
corresponds to filling in a dip with subsequent nucleation of a
growing tip.

Energy and momentum are associated with the noise-induced  left hand 
solitons moving on the noisy manifold $p=2\nu u$. By inspection of 
(\ref{3u}) it follows that the total energy $E=-(32/3)\nu\lambda u^3$,
the total momentum $\Pi =0$, and the total area $M=0$ are conserved 
during the collision.

Choosing the amplitude $u=2$ we have in Fig.~6 depicted the numerical
verification of the slope field $u$ as a function of $x$ for different
values of $t$. In Figs.~7 and 8 we have shown the associated noise
field $p$ and the height profile $h$ as a function of $x$ for the same
values of $t$. In Fig.~9 we have shown a gray-scale representation
of $u$ in the $xt$ plane. We notice that there is no phase shift
associated with the scattering process. Finally, in Fig.~10 we 
have shown the convergence of energy, momentum, and area during the
forward and backward time sweeps integrating numerically the field 
equations.
\begin{figure}
\begin{picture}(100,380)
\put(-10.0,40.0)
{
\centerline
{
\epsfxsize=7cm
\epsfbox{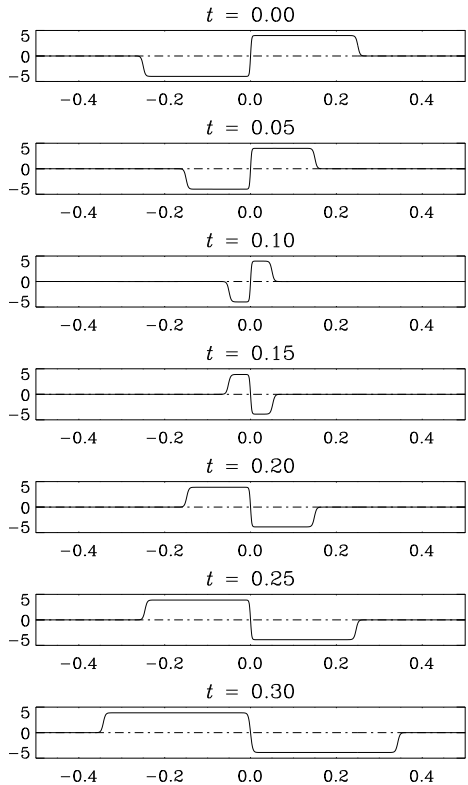}
}
}
\end{picture}
\caption{
Three soliton collision:
The slope field $u$ as a function of $x$ for different values of $t$.
}
\end{figure}
\begin{figure}
\begin{picture}(100,300)
\put(0.0,10.0)
{
\centerline
{
\epsfxsize=7cm
\epsfbox{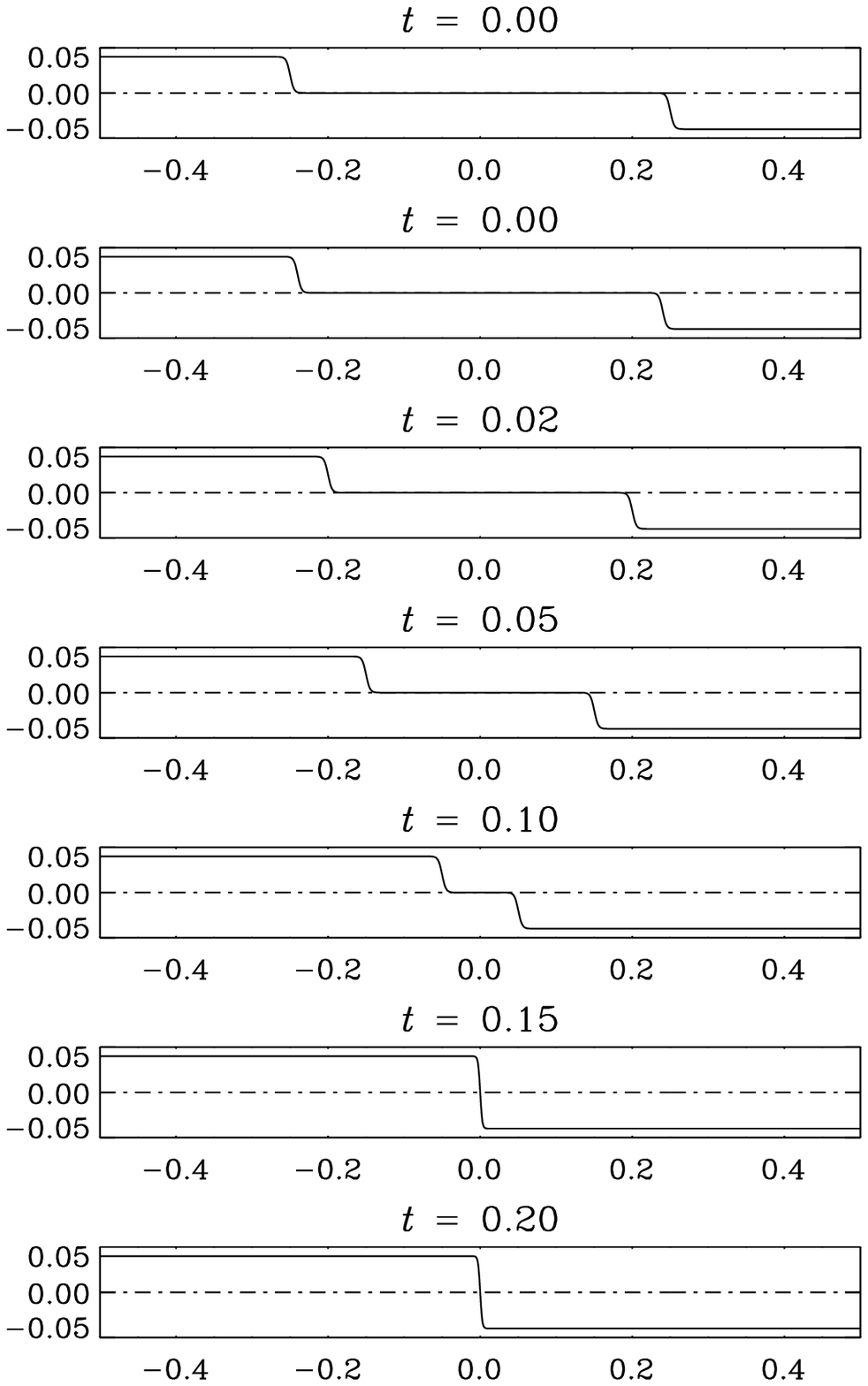}
}
}
\end{picture}
\caption{
The noise field $p$ as a function of $x$ for the same values of $t$
as in Fig.~6.
}
\end{figure}
\begin{figure}
\begin{picture}(100,220)
\put(0.0,10.0)
{
\centerline
{
\epsfxsize=7cm
\epsfbox{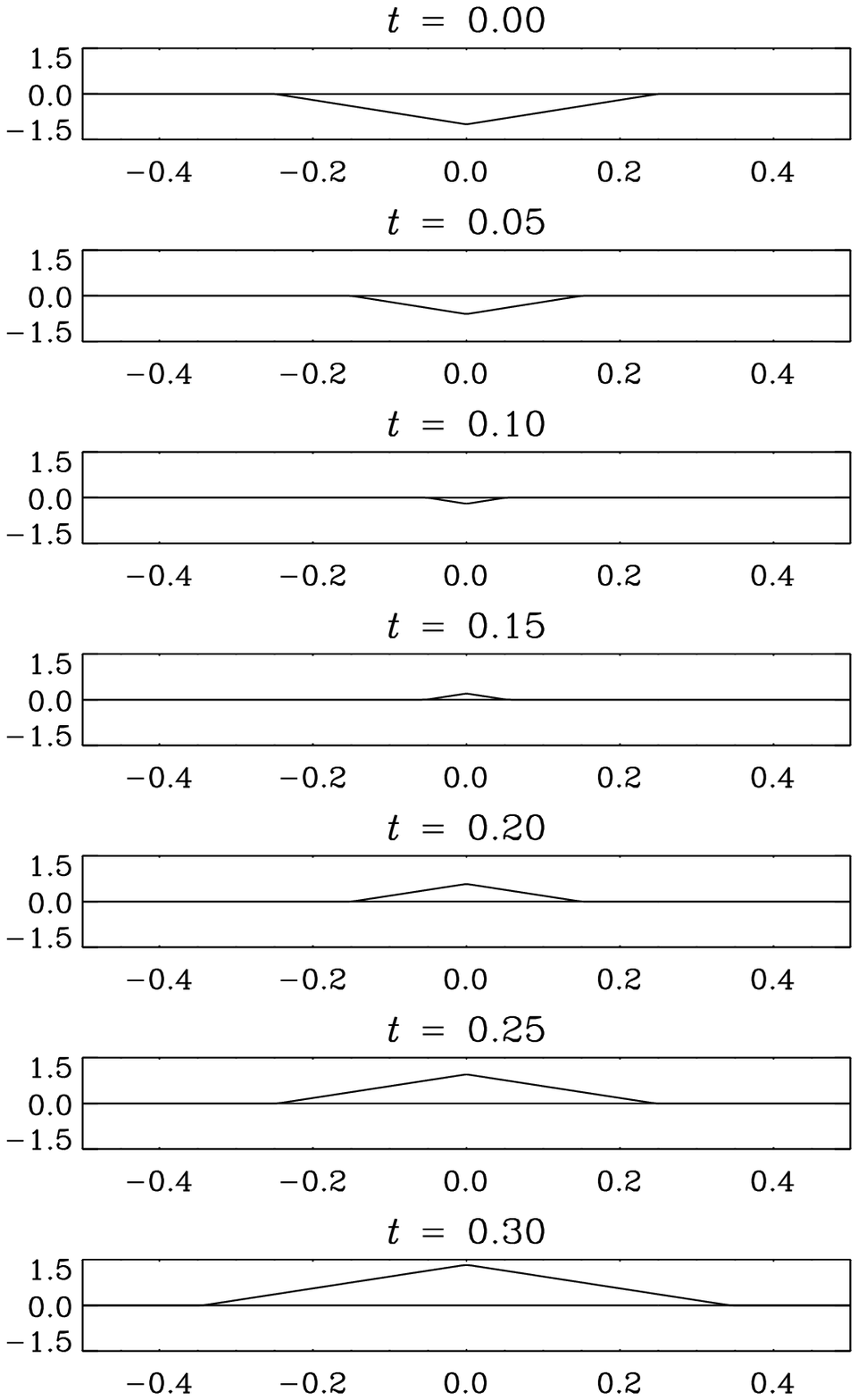}
}
}
\end{picture}
\caption{
The height profile $h=\int u\,{\rm d}x$ as a function 
of $x$ for the same values of $t$
as in Fig.~6.
}
\end{figure}
\begin{figure}
\begin{picture}(100,250)
\put(-10.0,20.0)
{
\centerline
{
\epsfxsize=6.5cm
\epsfbox{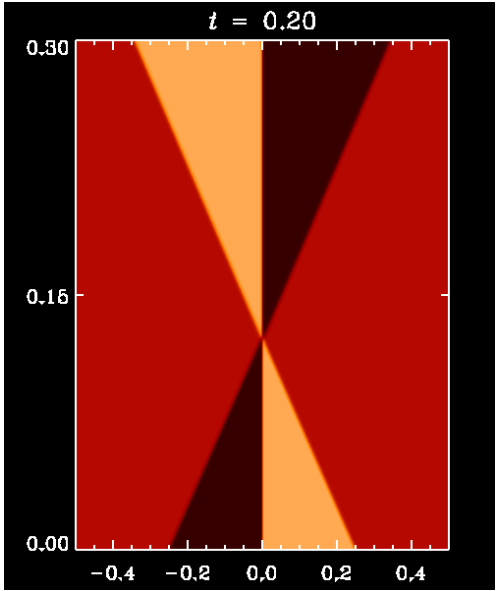}
}
}
\end{picture}
\caption{
Gray-scale representation of $u$ in the $xt$ plane showing the
three soliton collision. Note the
absence of a phase shift during the collision.
}
\end{figure}
\begin{figure}
\begin{picture}(100,270)
\put(-10.0,20.0)
{
\centerline
{
\epsfxsize=7cm
\epsfbox{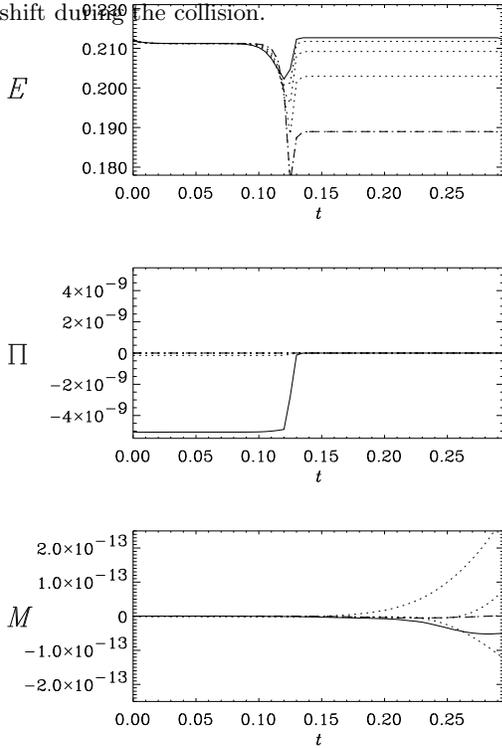}
}
}
\end{picture}
\caption{
Evolution of $E$, $\Pi$, and $M$ during the first few
iteration steps. The dashed line indicates the first
iteration step and the solid line the last one. Dotted
lines are for intermediate steps.
}
\end{figure}
\subsection{Pair soliton collisions}
In the second case we consider the collision of
two pair solitons of equal size and amplitude. 
The trial solution propagating with velocity $v=\lambda u$ has
the form
\begin{eqnarray}
u(x,t)=&&
-[u_1(x-vt+x_1)-u_1(x-vt+x_2)]
\nonumber
\\
&&
+[u_1(x+vt-x_2)-u_1(x+vt-x_1)]
\\
p(x,t)=&& 
-2\nu[u_1(x-vt+x_1)+u_1(x+vt-x_1)]
\end{eqnarray}
for $0<vt<x_2$,
\begin{eqnarray}
u(x,t)=&&
-[u_1(x-vt+x_1)-u_1(x)]
\nonumber
\\
&&
+[u_1(x)-u_1(x+vt-x_1)]
\\
p(x,t)=&&
-2\nu[u_1(x-vt+x_1)+u_1(x+vt-x_1)]
\end{eqnarray}
for $x_2<vt<x_1$,
\begin{eqnarray}
u(x,t)=&&
+[u_1(x+vt-x_1)-u_1(x)]
\nonumber
\\
&&
-[u_1(x)-u_1(x-vt+x_1)]
\\
p(x,t)=&& 
-2\nu[u_1(x)+u_1(x)]
\end{eqnarray}
for $x_1<vt<2x_1-x_2$, and
\begin{eqnarray}
u(x,t)=&&
+[u_1(x+vt-x_1)-u_1(x+vt-2x_1+x_2)]
\nonumber
\\
&&
-[u_1(x-vt+2x_1-x_2)-u_1(x-vt+x_1)]
\nonumber
\\
&&
\\
p(x,t) =&&
-2\nu[u_1(x+vt+2x_1-x_2)
\nonumber
\\
&&
+u_1(x-vt+2x_1-x_2)]
\end{eqnarray}
for $2x_1-x_2<vt$.

In this mode two pair solitons of amplitude $2u$ propagate with equal
and opposite velocities toward one another. The two leading kink
solitons merge to a static soliton and the two trailing kinks are
absorbed. Subsequently, the static right hand soliton flips over to a
static left hand soliton and the two pair solitons re-emerge.
Ana\-ly\-zing the collision it follows that the scattering of pair solitons
is transparent and accompanied by a phase shift in space equal to
the soliton size $|x_2-x_1|$ or, equivalently, a time delay
$|x_2-x_1|/v$. In terms of the associated height profile the scattering
situation corresponds to filling in a trough due to two colliding 
steps and the subsequent nucleation of a growing plateau.

By inspection it again follows that the total energy
$E=-(32/3)\nu\lambda u^3$, the total momentum $\Pi=0$, and the total
area $M=0$ are conserved during collision.

Choosing the amplitude $2u$ and the kink positions $x_1=0.25$
and $x_2=0.15$ we have shown in Fig.~11 the numerical verification
of the slope field $u$ as a function of $x$ for different values of
$t$. In Figs.~12 and 13 we have shown the associated noise field
$p$ and the height profile $h$ as a function of $x$ for the same 
values of $t$. In Fig.~14 we have shown a gray-scale representation 
of $u$ in the $xt$ plane. We notice the phase shift engendered during
the transparent collision. Finally, in Fig.~15 we have shown the
convergence of energy, momentum, and area during the forward and 
backward time sweeps in the numerical integration.
\begin{figure}
\begin{picture}(100,310)
\put(0.0,10.0)
{
\centerline
{
\epsfxsize=7cm
\epsfbox{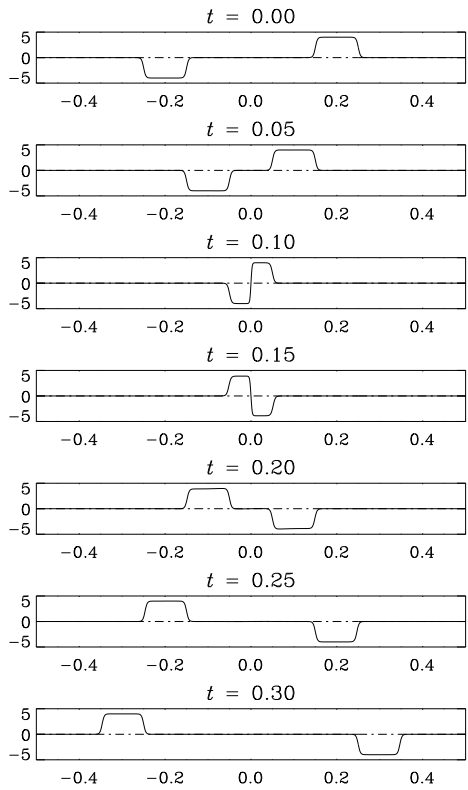}
}
}
\end{picture}
\caption{
Pair soliton collision:
The slope field $u$ as a function of $x$ for different values of $t$.
}
\end{figure}
\begin{figure}
\begin{picture}(100,220)
\put(0.0,10.0)
{
\centerline
{
\epsfxsize=7cm
\epsfbox{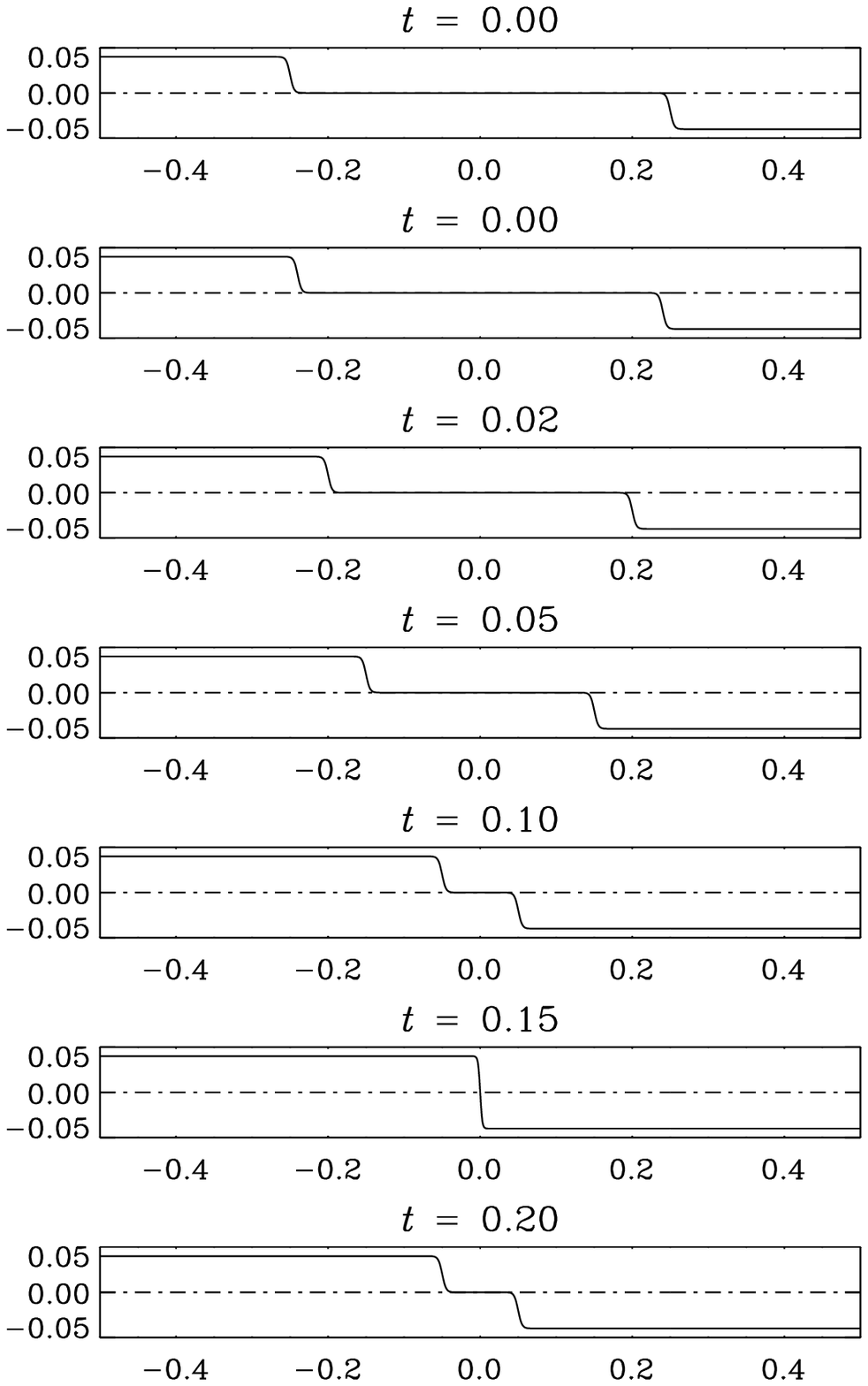}
}
}
\end{picture}
\caption{
The noise field $p$ as a function of $x$ for the same values of $t$
as in Fig.~11.
}
\end{figure}
\begin{figure}
\begin{picture}(100,310)
\put(0.0,10.0)
{
\centerline
{
\epsfxsize=7cm
\epsfbox{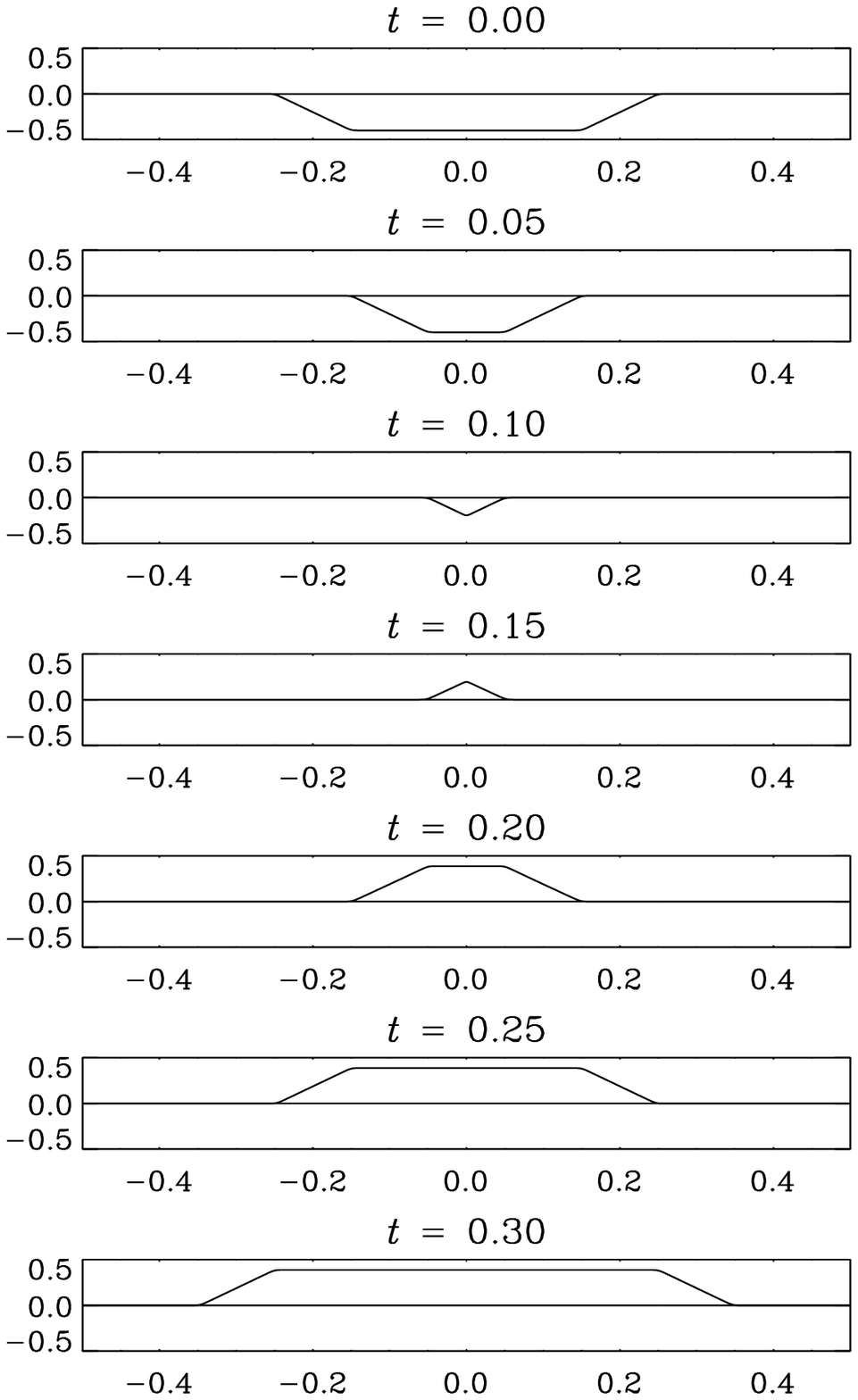}
}
}
\end{picture}
\caption{
The height profile $h=\int u\,{\rm d}x$ as a function 
of $x$ for the same values of $t$
as in Fig.~11.
}
\end{figure}
\begin{figure}
\begin{picture}(100,260)
\put(0.0,20.0)
{
\centerline
{
\epsfxsize=6.5cm
\epsfbox{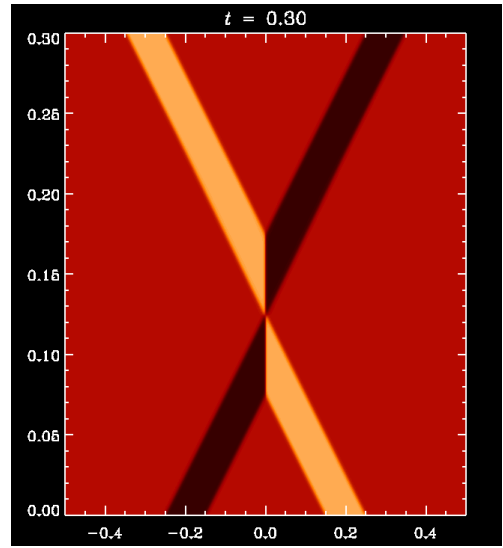}
}
}
\end{picture}
\caption{
Gray-scale representation of $u$ in the $xt$ plane showing the
pair soliton collision. Note the
occurrence of a phase shift during the collision.
}
\end{figure}
\begin{figure}
\begin{picture}(100,300)
\put(-10.0,0.0)
{
\centerline
{
\epsfxsize=7cm
\epsfbox{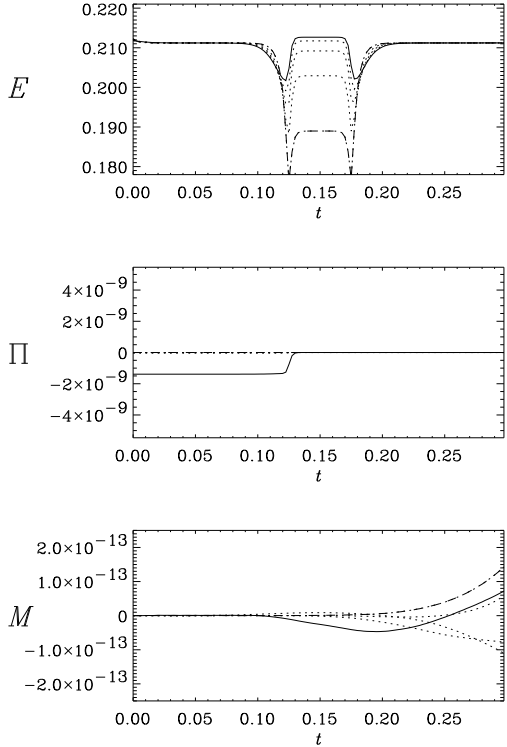}
}
}
\end{picture}
\caption{
Conserved quantities depicted as in Fig.~10.
}
\end{figure}

\section{Growth and nucleation}
Since we have achieved numerical justification of three
specific dynamical soliton configurations, namely i) the pair
soliton, ii) the collision of two solitons with a static
soliton, and iii) the collision of two pair solitons, we can
proceed to draw some simple conclusions based on the general
framework discussed in Sect.~1. There are two levels of description:
the stochastic Langevin level and the deterministic Fokker-Planck
or equation of motion level. On the Fokker-Planck level yielding
the canonical field equations (\ref{mfe1}) and (\ref{mfe2}) the
growth of the interface is interpreted in terms of a gas of 
propagating solitons (and diffusive modes). The stochastic
description on the Langevin level is then established in the
weak noise limit $\Delta\rightarrow 0$ by computing the action $S$
associated with a particular dynamical mode and subsequently deduce
the probability distribution according to (\ref{dis}), i.e.,
$P\propto\exp(-S/\Delta)$. This procedure is completely equivalent
to the WKB limit of quantum mechanics. Here the wavefunction 
$\Phi$ and thus the probabilistic interpretation is given by
$\Phi\propto\exp(iS/\hbar)$, where $S$ is the action associated
with the classical motion \cite{Landau59c}. Note that unlike quantum mechanics 
there is no phase interference in the stochastic nonequilibrium
case. 
\subsection{The pair soliton: anomalous diffusion}
The long-lived pair soliton (\ref{2sol}) has size 
$\ell = |x_1-x_2|$, amplitude $2u$, and propagates with velocity
$v=-\lambda u$. During a revolution in a system of size $L$ with
periodic boundary conditions the height field increases with a
layer of thickness $2u\ell$. Since the system is traversed in time
$t=L/v$ the integrated growth velocity is given by
$2\lambda u^2\ell/L$ which for a single pair of fixed size vanishes in the
thermodynamic limit. On the other hand,the local growth 
velocity $dh/dt$ is given by
$2\lambda u^2=(\lambda/2)(\nabla h)^2$ which is consistent with the
averaged KPZ equation (\ref{kpz}) in the stationary state.

The stochastic properties of the pair soliton growth mode is also
easily elucidated by noting that the action associated with the pair
mode is given by $S=(4/3)\nu\lambda|u|^3t$. Denoting the center
of mass of the pair mode by $x=(x_1+x_2)/2$ we have
$u=v/\lambda=x/t\lambda$ and we obtain using (\ref{dis}) the transition
probability
\begin{eqnarray}
P(x,t)\propto\exp\left(-\frac{4}{3}\frac{\nu}{\Delta\lambda^2}
\frac{x^3}{t^2}\right)~,
\label{rwpm}
\end{eqnarray}
for the `random walk' of independent pair solitons or steps in the
height profile. Comparing (\ref{rwpm}) with the distribution for
`ordinary' random walk originating from the Langevin equation
$dx/dt=\eta, \langle\eta\eta\rangle(t)=\Delta\delta(t)$,
$P(x,t)\propto\exp(-x^2/2\Delta t)$, we observe that the growth mode
performs anomalous diffusion. The distribution (\ref{rwpm}) also
implies the soliton mean square displacement, assuming pairs of
the same average size,
\begin{eqnarray}
\langle x^2\rangle(t)\propto
\left(\frac{\Delta\lambda^2}{\nu}\right)^{1/z}t^{2/z}~,
\label{msd}
\end{eqnarray}
with dynamic exponent $z=3/2$, identical to the dynamic exponent
defining the KPZ universality class. This result should be contrasted
with the mean square displacement 
$\langle x^2\rangle\propto\Delta t^{2/z}$,
$z=2$, for ordinary random walk. The growth modes thus perform
superdiffusion. 
\subsection{Soliton collisions: nucleation}
The three soliton scattering case shown in Figs.~6, 7, and 8
corresponds to the filling in of a dip and the subsequent
nucleation of a tip. The incoming solitons have amplitude $2u$
and move with velocities $v=\pm\lambda u$. Denoting the distance
of a soliton from the center by $d$, the height change from
the bottom of the dip to the top of the tip is $\Delta h = 4ud$.
The duration of the collision process is $\Delta t=2d/v$ and we
obtain for the local growth velocity 
$\Delta h/\Delta t=(\lambda/2)(2u)^2=(\lambda/2)(\nabla h)^2$,
consistent with the KPZ equation.

As regards the stochastic properties the action prior to 
the collision is associated with the incoming left hand solitons
and given by $(8/3)\nu\lambda u^3 t$. After the collision the
static left hand soliton of amplitude $4u$ carries the action
$(32/3)\nu\lambda u^3 t$. There is thus an increase of action
in connection with the tip formation yielding, according to
(\ref{dis}), a reduced probability, i.e., a statistical suppression
of tip nucleation.

The case of pair soliton scattering shown in Figs.~11, 12, and 13
correspond to the filling in of a trough and the subsequent formation
of a growing plateau. The incoming pair solitons have amplitude $2u$ and
propagate with velocities $v=\pm\lambda u$. Denoting the pair size by
$d$ the height change from the trough to the plateau is 
$\Delta h = 4ud$, the duration of the collision $\Delta t = 2d/v$,
and we obtain again $\Delta h/\Delta t = (\lambda/2)(\nabla h)$,
consistent with the KPZ equation.

Before and after the collision the action is now unchanged and equal to
$(8/3)\nu\lambda u^3 t$. As in the three soliton case this action
comes from the two left hand solitons. However, during the
collision in a time interval $\Delta t = d/v$ the action jumps
to the value $(32/3)\nu\lambda u^3 t$ as the static left hand soliton
of amplitude $4u$ is temporarily formed before the re-emergence 
of the pair solitons. Comparing this situation with the tip formation
in the three soliton scattering case we conclude that tip formation
is statistically suppressed with respect to the formation of a 
growing plateau. Generally the formation of tips is `expensive'
since a large change in slope yields a large change in action.
\section{Summary and conclusion}
In the present paper we have numerically investigated the coupled
diffusion-advection type field equation originating from 
the canonical phase space approach applied to the noisy
Burgers equation or the equivalent KPZ equation in one spatial
dimension. We have shown that the pair soliton mode in the 
slope field corresponding to moving steps in the height field
forms a long-lived excitation. We have furthermore investigated 
two special scattering scenarios, namely the collision of two
identical moving solitons with a static soliton and the collision
of two identical pair solitons.
They correspond in the height field respectively to the
nucleation of a growing tip and to the formation of a growing plateau.
Finally, we have applied the canonical phase space approach in order
to estimate the stochastic aspects of the above configurations
and found i) that a step in the height field performs a random walk
with dynamical exponent $z=3/2$ corresponding to averaged superdiffusion
and ii) that tip formation is stochastically suppressed with
respect to plateau formation.

As discussed above the inherently unstable structure of the 
field equation makes a direct
integration forward in time inaccessible and we can thus not establish
solutions as an initial value problem and discuss the equations
generically. Consequently, we are limited to numerically check 
trial solutions representing a variety of scattering situations.
So far we have only been able to verify the symmetric cases of three soliton
and soliton pair collisions. In order to extend the present numerical approach
and thus provide substance to the heuristic quasi-particle
representation of a growing interface it is clearly of interest
to design more involved trial solutions. Alternatively, a completely
different approach to generate solutions is called for.

\end{multicols}
\bibliography{articles,books}

\end{document}